\documentclass[aps,pra,preprint,amsmath,amssymb,showpacs,preprintnumbers]{revtex4}

\usepackage{dcolumn}
\usepackage{graphicx}
\usepackage{bm}
\input epsf

\begin{document}

\bibliographystyle{unsrt}

\preprint{}

\title{Berry-like phases in structured atoms and molecules}
\author{Edmund R. Meyer$^1$}
\email{meyere@murphy.colorado.edu}
\author{Aaron Leanhardt$^2$}
\author{Eric Cornell$^1$}
\author{John L. Bohn$^1$}
\affiliation{${}^1$JILA, NIST and University of Colorado, Department of Physics,
  Boulder, Colorado 80309-0440, USA}
\affiliation{${}^2$Department of Physics, University of Michigan, Ann Arbor, 
  Michigan 48109-1040, USA}
\date{\today}

\begin{abstract}
Quantum mechanical phases arising from a periodically varying Hamiltonian 
are considered.  These phases are derived from the eigenvalues of a
stationary, ``dressed'' Hamiltonian that is able to treat internal atomic 
or molecular structure in addition to the time variation. In the limit of 
an adiabatic time variation, the usual Berry phase is recovered.  
For more rapid variation, non-adiabatic corrections to the Berry phase are 
recovered in perturbation theory, and their explicit dependence on internal 
structure emerges. Simple demonstrations of this formalism are given, to 
particles containing interacting spins, and to molecules in electric fields.
\end{abstract}


\maketitle

%%%%%%%%%%%%%%%%%%%%%%%%%%%%%%%%%%%%%%%%%%%%%%%%%%%%%%
\section{Introduction}

Any quantum mechanical system in a stationary state accumulates a
dynamical phase over time proportional to the energy of that state.
To determine energy differences, based on phase differences
accumulated between two such states, is the basis of Ramsey
spectroscopy; the workhorse of high-precision measurement.  For this
reason, small effects that can add spurious phase shifts must be
understood and kept under control.

One such variety of spurious phases arises when the states of
interest are not strictly stationary. If the total Hamiltonian
$H(t)$ has an explicit time dependence, then this dependence will
generate an additional phase evolution.  For example, precision
spectroscopy of trapped ions must contend with the fact that the
ions are in motion, and experience varying ambient fields
during the course of their orbit.  Likewise, ultracold atomic
samples confined in an optical dipole trap experience, in principle,
an oscillating electric field due to the laser field that provides
the trapping potential.

In the case of a Hamiltonian with a slow, periodic time dependence
$H(t+\tau) = H(t)$, Berry \cite{berry84} has given a famous description of the
additional phase.  Berry's original treatment requires that the
period $\tau$ be far larger than any other relevant time scale of
the system, and thus finds an ``adiabatic'' phase shift.  This shift
is largely independent of the detailed way in which the Hamiltonian
varies with time, and leads to an elegant geometric description of
the phase \cite{berry84,simon83,anandan87,aharonov87,page87,garrison,robbins94}.
Extensions to this formalism have considered the next-order corrections if
the rate of change of the Hamiltonian is not strictly adiabatic
\cite{berry87,berry90,berry90-2,cui92,hannay98,moore90-2,moore91-2}.
A more general Floquet theory has also been advanced, which allows one to
consider the effect of overtones of the fundamental period $\tau$
\cite{moore90,moore91,moore91-2}. In addition, the ideas have been extended
to particles with dynamic properties \cite{horsley07}, gauge
structure \cite{wilczek84}, the Quantum Hall Effect \cite{yaffe87}, and to 
relativistic effects using the Dirac equation \cite{wang99}.

\begin{figure}\label{fieldfig}
  \begin{center}
    \resizebox{4in}{!}{\includegraphics{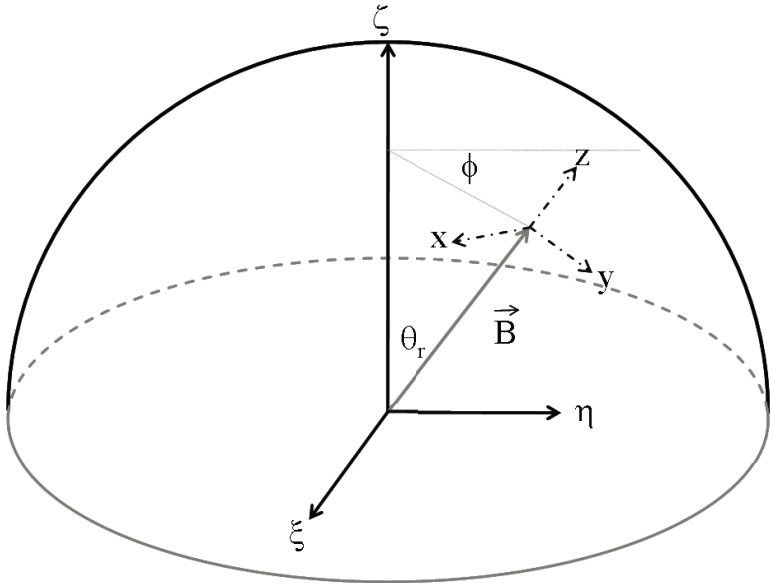}}
  \end{center}
  \caption{The axis of rotation with laboratory-fixed coordinates 
    $\{\xi,\eta,\zeta\}$ as well as the field coordinates defined by
    $\{x,y,z\}$. The field direction rotates about the $\zeta$-axis 
    with angular frequency $\omega_r$.}
\end{figure}

Thus far, applications of the Berry phase have mostly considered the
effect of the time-dependence on quantum mechanical particles
without internal structure, although atoms with two or several levels
have been considered \cite{hoodbhoy88,moore91,DeMille09}. 
However, the job of precision spectroscopy is precisely to reveal this 
internal structure. Corrections to Berry's phase arising from degrees of 
freedom internal to an atom or molecule is our concern in this article.  
To establish a concrete formalism for this, we will consider a
particular case, namely, a diamagnetic or electrically polar species
in the presence of a magnetic or electric field, whose direction precesses
on a cone with an angular frequency $\omega_r$ (see Fig.\,\ref{fieldfig}).
The system evolves in time according to the field variation, combined
with whatever intrinsic Hamiltonian governs the particle's internal structure. 
The internal structure dictates regimes of linear and quadratic Zeeman 
(Stark) shifts with respect to the applied magnetic (electric) field.

A main point in deriving the non-dynamic phase in this situation is
to recognize the periodicity of the driving field.  By analogy to
the periodic driving of a near-resonant laser field applied to a
two-level atom, we consider ``field-dressed'' states of the
Hamiltonian in the spirit of quantum optics \cite{cohen,DeMille09}.  
This viewpoint effectively counts the energy of the atom itself, plus that 
of the photons of frequency $\omega_r$ arising from the driving field. The 
additional energy shift due to the rotating field is then equivalent to the 
ac Stark effect in optics. By constructing the complete Hamiltonian in this 
way, we are able to accommodate the particle's internal structure. We are also 
able to consider arbitrary rates of rotation, not just those that are
adiabatic with respect to the particle's Hamiltonian. Nevertheless, in 
what follows we will focus primarily on results for low rotation rates, to
better draw analogies with the usual adiabatic phase.

This article is structured as follows.  In Sec.\,II we work out the
general transformation from a Hamiltonian with an explicit
time-rotating field, into an effective dressed Hamiltonian whose
eigen-energies yield the shifted energies.  We illustrate this
transformation first with a simple two-level atom, then generalize
it to an arbitrary atom or molecule. In Sec.\,III we briefly
re-visit a structureless particle with arbitrary total spin, showing
that our results reduce to Berry's in the limit of slow rotation.
Sec.\,IV illustrates the application of the method to a particle
composed of two interacting spin-1/2 objects, to show most clearly
the effect of their fine structure on the Berry phase.  Finally, Sec.\,V
considers a simple case of a dipolar molecule in a rotating electric
field, to assess the influence of molecular end-over-end rotation on
the phase.

%%%%%%%%%%%%%%%%%%%%%%%%%%%%%%%%%%%%%%%%%%%%%%%%%%%%%%
\section{Derivation of the dressed states}

This section lays the ground work for the formalism. It is composed
of two subsections. The first works out the derivation for a
spin$-1/2$ system in a rotating magnetic field, and will illustrate
simply and cleanly the basic idea.  The second subsection contains a
generalization of this same derivation to explicitly account for an
arbitrary quantum mechanical object with total spin $j$.

\subsection{Spin$-1/2$ interaction}

The most elementary of quantum mechanical objects is the spin$-1/2$
particle. Its interaction with a time-dependent magnetic field is
\begin{eqnarray}
  H(t) = -{\vec \mu}_m \cdot {\cal {\vec B}}(t).
\end{eqnarray}
There are two coordinate systems that we will find useful, the
lab-fixed Cartesian coordinates labeled $(\xi, \eta, \zeta)$; and
a coordinate system $(x,y,z)$ rotating with the magnetic field,
where ${\hat z} = {\cal {\hat B}}$ (see Fig.\,\ref{fieldfig}). In the 
lab-fixed frame the field rotation is given by its components
\begin{eqnarray}\nonumber
  {\cal {\vec B}}(t) &=& \left( {\cal B}_{\xi}(t), {\cal B}_{\eta}(t), 
  {\cal B}_{\zeta}(t) \right) \\
  &=& {\cal B} \left( \sin \theta_r \cos \omega_r t, \sin \theta_r \sin
  \omega_r t, \cos \theta_r \right),
\end{eqnarray}
where $\phi = \omega_r t$. The field makes an angle $\theta_r$ with respect
to the lab-fixed $\zeta$ axis, which is also the axis it rotates about.
In other words we identify a vector ${\vec \omega}_r = \omega_r {\hat\zeta}$,
the axis defining the field's rotation.

The spin-field interaction has its own characteristic frequency,
\begin{eqnarray}
  -{\vec \mu}_m \cdot {\cal {\vec B}} = \frac{g \mu_B {\cal B}}{2} {\vec \sigma}
  \cdot {\cal {\hat B}} \equiv \frac{\hbar}{2} \omega_L {\vec \sigma} \cdot {\hat B},
\end{eqnarray}
in terms of the Pauli matrices ${\vec \sigma}$ and the ($m$-independent) Larmor
frequency $\omega_L = g\mu_B {\cal B}$.  $g$ is the usual $g$-factor; e.g. 
$g\approx 2.0023$ for the simple case of the spin-$1/2$ electron. The 
eigen-energies of the non-rotating spin-field interaction are 
$\pm \hbar\;\omega_L/2$.  By causing the field to rotate, we expect to find 
apparent energies that are different from these. From this point forward we 
shall work in atomic units and set $\hbar = 1$.

In this simple example, we can easily cast the Hamiltonian in matrix
form, using the basis of spin functions $|m_\zeta = \pm 1/2 \rangle$, with
quantization along the $\zeta-$axis.  These states themselves are
then independent of time, which would not be the case if we used the
eigenstates referred to the (moving) magnetic field. Using the
explicit forms for the Pauli matrices, we get
\begin{equation}
  H(t) = \frac{\omega_L}{2} \left( \begin{array}{cc} \cos
    (\theta_r) & \sin (\theta_r)\; e^{-i \omega_r t} \\  \sin (\theta_r)\;
    e^{i \omega_r t} & -\cos (\theta_r) \end{array} \right).
\end{equation}
While this basis seems cumbersome it is quite useful for the
dressed-state calculation we wish to perform.

We begin in the most general way by writing an {\it ansatz} wave
function in this basis:
\begin{eqnarray}\label{spin_half_ansatz}
  |\psi(t) \rangle = \left( \begin{array}{c} \alpha(t) e^{-i \omega_+
      t} \\ \beta(t) e^{-i \omega_- t} \end{array} \right),
\end{eqnarray}
where we leave the frequencies $\omega_{\pm}$ unspecified for now.
We note a very similar development has recently been carried out by the authors
of Ref.\,\cite{qing09} for a two-level system. The constants  will be
chosen later to cancel any time dependence in the system. The wave function
$|\psi\rangle$ satisfies the time-dependent Schr\"{o}dinger equation (TDSE)
\begin{eqnarray}
  i {\partial |\psi \rangle \over \partial t} = H(t) |\psi \rangle.
\end{eqnarray}
Inserting the {\it ansatz} \eqref{spin_half_ansatz} into the TDSE yields the
equation of motion for $\alpha$ and $\beta$. This new equation introduces
terms which amount to a kind of effective Hamiltonian due to rotation. Moving
these to the RHS of the TDSE, we get
\begin{eqnarray}
  i {\partial \over \partial t} \left( \begin{array}{c} \alpha \\
    \beta
  \end{array} \right)= \left( \begin{array}{cc}  + \frac{\omega_L}{2}
    \cos \theta_r -\omega_+ & \frac{\omega_L}{2} \sin \theta_r e^{-i(\omega_r +
      \omega_- - \omega_+)t} \\ \frac{\omega_L}{2} \sin \theta_r e^{i(\omega_r +
      \omega_- - \omega_+)t} & -\frac{\omega_L}{2} \cos \theta_r - \omega_-
  \end{array} \right).
\end{eqnarray}
Formally, this is still a time-dependent Hamiltonian unless we use
the freedom in choosing the frequencies $\omega_{\pm}$ to get rid of
this dependence.  We could achieve this by setting
\begin{eqnarray}
  \omega_r + \omega_- = \omega_+.
\end{eqnarray}
There are many ways to accomplish this, but a particularly symmetric
and appealing one is to let
\begin{eqnarray}
  \omega_{\pm} = \pm {\omega_r \over 2},
\end{eqnarray}
or, even more to the point,
\begin{eqnarray}\label{awesome_guess}
  \omega_{m_\zeta} = m_\zeta\; \omega_r
\end{eqnarray}
for each angular momentum projection $m_\zeta$ along the rotation axis.
This last statement is very general, and will motivate our choice of
wave function in the spin-$j$ case below.

In any event, we are presented with a formally time-independent
Hamiltonian, dressed by the rotation:
\begin{eqnarray}\label{dressed_Hamiltonian}
  H_{\rm dressed} = H_{\rm nr} + H_{\rm r},
\end{eqnarray}
where the magnetic-field part, in the laboratory-fixed $|m_\zeta\rangle$ basis,
looks like
\begin{eqnarray}
  H_{\rm nr} = \frac{\omega_L}{2}\left( \begin{array}{cc} \cos (\theta_r) &
    \sin (\theta_r) \\ \sin (\theta_r) & -\cos (\theta_r)
\end{array} \right),
\end{eqnarray}
and is exactly the same as having the magnetic field tilted at an
angle $\theta_r$ with respect to the quantization axis in the
direction of the $x$-axis, as specified by $\phi = \omega_r t = 0$.
Rather, the vestiges of rotation show up in the effective term
\begin{eqnarray}
  H_{\rm r} = \frac{\omega_r}{2}\left( \begin{array}{cc} -1 & 0 \\ 0 & +1
  \end{array} \right) = -{\rm diag} ( m_\zeta\;\omega_r).
\end{eqnarray}

%%%%%%%%%%%%%%%%%%%%%%%%%%%%%%%%%%%%%%%%%%%%%%%%%%%%%%%%%%

Now that we have a time-independent Hamiltonian we write
\begin{eqnarray}
  \left( \begin{array}{c} \alpha(t) \\ \beta(t) \end{array} \right) = \left(
  \begin{array}{c} \alpha_0 \\ \beta_0 \end{array} \right) \exp ( -i
  \lambda t),
\end{eqnarray}
where $\lambda$ is the effective, or dressed, eigen-frequency of
the ``stationary'' state defined by $H_{\rm nr} + H_{\rm r}$.
$\lambda$ is, in other words, the eigenvalue of $H_{\rm dressed}$:
\begin{equation}\label{spin_half_lambdas}
  \lambda_{\pm}  = \pm \frac{1}{2}\sqrt{ \omega_L^2 - 2\omega_L \omega_r \cos
    (\theta_r) + \omega_r^2}.
\end{equation}
Thus the apparent energy difference, measured by the phase
difference accumulated during a hold time $\tau$, would be
$(\lambda_+ - \lambda_-)\tau = \phi_+ - \phi_-$, rather than the 
$(\omega_{+1/2} - \omega_{-1/2}) \tau = \varphi_+ - \varphi_-$ that 
would measure the intrinsic energy splitting. The difference between 
these two we attribute to a Berry-like phase in each state defined as 
(where we assume the dynamical phase is unaffected by the rotation)
\begin{equation}\label{gamma_def}
  \gamma_\pm = \phi_\pm - \varphi_\pm,
\end{equation}
where $\phi$ ($\varphi$) is the total (dynamic) phase of the system. 
This removes the dynamical phase $\varphi$ that would have
been accumulated in the absence of rotation.  Hence $\gamma/\tau$ represents
the error introduced by the field's rotation into a measurement of
the state's energy.

The dressed eigen-energies are actually independent of whether or not
the time $\tau$ refers to one period of the rotating field.  To connect
explicitly to Berry's phase, let $\tau = 2 \pi / \omega_r$ be one period
of the field's rotation. Berry's phase results when the field rotation
is slow, i.e. when the frequency of the rotating field is small compared
to the Larmor precession frequency of the spin. If we assert that
$\omega_r \ll \omega_L$, then the spin is
expected to precess around the instantaneous field direction, and to
follow this direction during the rotation of the field.  In this
limit the dressed-state eigenvalue of one state, say $\lambda_+$, is
\begin{eqnarray}
  \lambda_+ \approx \frac{\omega_L}{2} - \frac{\omega_r}{2}\cos (\theta_r).
\end{eqnarray}
In a period $\tau$ the additional phase accumulated by this state is
(see \eqref{gamma_def})
\begin{eqnarray}\label{spin_half_phase}
  \gamma_+^{(0)} & \approx & - \frac{1}{2} \omega_r \tau \cos
  (\theta_r) \nonumber \\ & \approx & \pi - \frac{1}{2} (2 \pi) \cos (\theta_r).
\end{eqnarray}
Here we have used the freedom to add $\pi$ to the phase, which 
amounts to an unobservable overall change of sign of the dressed
eigenstate $(\alpha_0 \; \beta_0)$.

The ordinary dynamical phase $\varphi=\omega_L \tau$ is canceled out in
\eqref{spin_half_phase} leaving only the terms due to the slow rotation
of the field. This remainder can be written
\begin{eqnarray}
  \gamma^{(0)} = \frac{1}{2} (2 \pi ) (1 - \cos (\theta_r)) = \frac{1}{2}
  \Delta \Omega \Rightarrow m\Delta\Omega,
\end{eqnarray}
where
\begin{eqnarray}
  \Delta \Omega = \int_0^{2 \pi} d \phi \int_0^{\theta_r} \sin
  (\theta) d\theta  = 2 \pi (1 - \cos (\theta_r))
\end{eqnarray}
is the solid angle subtended by the vector describing the direction
of the rotating field.  In other words, this additional phase 
$\gamma^{(0)}$ is exactly Berry's phase $m \Delta \Omega$, for a 
particle with spin projection $m = 1/2$ along the ${\hat \zeta}$-axis. 
Similarly, for the $m = -1/2$ state there arises an additional phase 
$-(1/2) \Delta \Omega$.

%%%%%%%%%%%%%%%%%%%%%%%%%%%%%%%%%%%%%%%%%%%%%%%%%%%%%%%%%%%%%%

We have cast the Hamiltonian $H$ in \eqref{dressed_Hamiltonian} in
the basis where angular momentum projection along the rotation axis
${\hat \omega}_r = {\hat \zeta}$ is a good quantum number.   Yet, in
the limit of slow rotation that is particularly important, the
relevant states are those where $m$ is quantized along the field
axis.  We should therefore cast $H$ in the basis of states $|s\,m
\rangle$ quantized along the field axis, which we call the ${\hat
z}$-axis (see Fig. 1).

Using the {\it ansatz} \eqref{spin_half_ansatz}, our dressed Hamiltonian
represents a Hamiltonian $H_{\rm nr}$ that is already diagonal in this basis,
and reads
\begin{eqnarray}
  H_{\rm nr} = \frac{\omega_L}{2}\left( \begin{array}{cc}1 & 0 \\ 0 &
    -1 \end{array} \right)
\end{eqnarray}
To cast the rotational piece $H_r$ in this basis, we make explicit
the notation $|s\; m_{\zeta} \rangle$ for spin states quantized along
the rotation axis ${\hat \zeta}$; and $|s\; m \rangle$ for spin states
quantized along the field axis ${\hat z}$.  The rotation from ${\hat
\zeta}$ to ${\hat z}$ is accomplished through the Euler angles $R =
(0, \theta_r, 0)$.  The transformation from $|s\; m_{\zeta} \rangle$
to $|s\; m \rangle$ is given by~\cite{bns}
\begin{eqnarray}
  |s\;m \rangle &=& D(R) |s\; m_{\zeta} \rangle \\
  &=& \sum_{m_{\zeta}^{\prime}} | s\;m_{\zeta}^{\prime} \rangle \langle
  s\;m_{\zeta}^{\prime} | D(R) | s\;m_{\zeta} \rangle \nonumber \\
  &=& \sum_{m_{\zeta}^{\prime}} | s\;m_{\zeta}^{\prime} \rangle
  D^s_{m_{\zeta}^{\prime} m_{\zeta}}(R).
\end{eqnarray}
The Wigner rotation matrices have a simple explicit form:
\begin{eqnarray}
D^s_{m_{\zeta}^{\prime} m_{\zeta}}(0, \theta_r, 0) =
e^{-im_{\zeta}^{\prime}0} d^s_{m_{\zeta}^{\prime}
m_{\zeta}}(\theta_r) e^{-im_{\zeta}0},
\end{eqnarray}
where
\begin{eqnarray}
  && d^{1/2}_{1/2, 1/2} = d^{1/2}_{-1/2, -1/2} = \cos \left({\theta_r \over 2}
  \right)  \\
  && d^{1/2}_{-1/2, 1/2} = -d^{1/2}_{1/2, -1/2} = \sin \left({\theta_r
    \over2}\right).
\end{eqnarray}
Using the fact that $H_{\rm r}$ is diagonal in the rotation basis with values
$-m_\zeta \omega_r$, it is easily verified that, in the rotating-frame basis,
\begin{equation}\label{Hrot_field}
  H_r = {\omega_r \over 2} \left( \begin{array}{cc} -\cos (\theta_r) &
    -\sin (\theta_r) \\ -\sin (\theta_r) & \cos (\theta_r)
  \end{array} \right).
\end{equation}

Therefore, the Hamiltonian matrix referred to the field axis is
\begin{eqnarray}
  H &=& H_{\rm nr} + H_{\rm r} \\
  &=& \left( \begin{array}{cc} \frac{\omega_L}{2} - \frac{\omega_r}{2}
    \cos \theta_r & -\frac{\omega_r}{2} \sin \theta_r \\ -\frac{\omega_r}{2}
    \sin \theta_r & -\frac{\omega_L}{2} + \frac{\omega_r}{2} \cos\theta_r
  \end{array} \right).
\end{eqnarray}
And this matrix gives has the same eigenvalues as (\ref{spin_half_lambdas}),
as is expected when one merely performs a unitary transformation on the system.

In this basis, the Hamiltonian is already diagonal in the absence of
rotation.  Thus for a small rotation it is nearly diagonal, and the
eigenvalues are easily estimated in perturbation theory.   Indeed,
Berry's energy follows immediately from the diagonal perturbations in
this matrix: it is $- (\omega_r /2 )\cos (\theta_r)$ for $m = +1/2$,
and $+ (\omega_r /2 )\cos (\theta_r)$ for $m = -1/2$.

To summarize, for a Hamiltonian of the form $H(t)$ that we have been
dealing with,  the rotation-dressed energies are given by the
eigenvalues of the time-independent operator
\begin{equation}
  H_{\rm dressed}=H(t=0) - \omega_r s_{\zeta} = H(t=0) - \omega_r \left( \cos
  (\theta_r) s_z - \sin (\theta_r) s_x \right).
\end{equation}
For more general systems incorporating internal structure, all that
is required is to include the appropriate structure in $H(t=0)$, as
we will now see.

%%%%%%%%%%%%%%%%%%%%%%%%%%%%%%%%%%%%%%%%%%%%%%%%%%%%%%%%%%%

\subsection{A general spin-$j$ system}

Having provided the groundwork by working out the simple spin-$1/2$
system, we now proceed with a general derivation for the spin-$j$
system. A system such as this can be described by the following
Hamiltonian
\begin{eqnarray}
  H(t) = H_0 - {\vec \mu} \cdot {\vec {\cal F}}(t),
\end{eqnarray}
where ${\vec {\cal F}}(t)$ is an external field, electric or magnetic,
that acts on an appropriate moment ${\vec \mu}$ of the atom or molecule.
${\vec {\cal F}}$ rotates on a cone at frequency $\omega_r$ and tilt
angle $\theta_r$ just as ${\vec B}$ did in the previous section and 
depicted in Fig.\,\ref{fieldfig}. Here $H_0$ is a Hamiltonian in the 
absence of the applied rotating field. It can be used to describe the 
hyperfine elements of an atom or it can be a detailed molecular Hamiltonian 
that includes such items as rotation, spin-spin, nuclear spin, or lambda 
doubling. The Hamiltonian can equally be represented in a basis referred to 
the axis of rotation or to the instantaneous field axis. Later we will take 
the instantaneous field axis, just as we did above. We will, as in the 
previous section, begin by quantizing along the axis of rotation. We 
take this structured object and place it in a rotating field 
${\vec {\cal F}}$; this can be either electric or magnetic provided there 
is an electric or magnetic dipole that interacts with the field in the 
usual way, i.e. it is a scalar interaction of two vectors.

To work with this Hamiltonian, it is again convenient to pick two
basis sets:
\begin{eqnarray}
  && | (\kappa ) j m_{\zeta} \rangle \nonumber \\
  && | (\kappa ) j m_j \rangle
\end{eqnarray}
Because $j$ is the total of all relevant angular momenta, its projection
onto an axis is unambiguously defined as $m_{\zeta}$ in the lab fame
and $m_j$ in the rotating frame, as above.  Here $\kappa$ is a shorthand
notation for all the other quantum numbers required to specify the state.

To deal with the explicit time dependence of the field rotation, we
will expand into the lab basis first, and will make the same
trial wave function that was motivated above;
\begin{eqnarray}
  | \psi (t) \rangle = \sum_{\kappa^{\prime}, j^{\prime} ,
    m_{\zeta}^{\prime} } C_{\kappa^{\prime}, j^{\prime} ,
    m_{\zeta}^{\prime} } e^{-i m_{\zeta}^{\prime} \omega_r t} |
  (\kappa^{\prime}) j^{\prime} m_{\zeta}^{\prime} \rangle.
\end{eqnarray}
We have explicitly included a time dependent phase factor with phase
$m_{\zeta} \omega_r$, which is akin to the spin-$1/2$ case, cf. 
\eqref{awesome_guess}. Taking the time derivative for the TDSE and 
projecting onto a particular state, gives
\begin{eqnarray}
  \langle (\kappa) j m_{\zeta}| i { d |\psi \rangle \over dt} = \left( i
      {\dot C}_{\kappa, j , m_{\zeta} } + m_{\zeta} \omega_r C_{\kappa, j ,
        m_{\zeta}} \right) e^{-i m_{\zeta} \omega_r t} .
\end{eqnarray}
As for the internal Hamiltonian $H_0$, it may or may not be diagonal
in our basis, but it does not depend on any external field.
Therefore it can be represented in a basis where it is diagonal 
in $m_{\zeta}$, whereby
\begin{eqnarray}
  \langle (\kappa ) j m_{\zeta} | H_0 | \psi \rangle =
  \sum_{\kappa^{\prime}, j^{\prime} , m_{\zeta}^{\prime} }
  e^{i (m_\zeta - m_{\zeta}^{\prime})\omega_r t}
  \langle (\kappa ) j m_{\zeta} | H_0 |
  (\kappa^{\prime}) j^{\prime} m_{\zeta}^{\prime} \rangle
  C_{\kappa^{\prime}, j^{\prime} , m_{\zeta}^{\prime} }
  \delta_{m_\zeta,m_\zeta^\prime}
\end{eqnarray}

To treat the field interaction, we use the language of tensor
algebra, and express the spherical components of ${\vec {\cal F}}$
in the lab frame as an explicit rotation from ${\vec {\cal F}}$ in
the rotating frame (whose $z$ axis is, of course, defined by the
instantaneous direction of ${\vec {\cal F}}$ itself):
\begin{eqnarray}
  {\cal F}_\iota  &=& \sum_{\iota}{\cal F}_q {\cal D}_{\iota q}^{1\star}
  (\omega_r t,\theta_r,0)\nonumber\\
  &=& {\cal F} {\cal D}_{\iota q}^{1\star}(\omega_r t,\theta_r,0).
\end{eqnarray}
${\cal F}$ is the magnitude of the field, and $q$ is its
spherical projection in the rotating frame. But ${\vec{\cal F}}$
defines this frame so only the values of $q=0$ will contribute.
${\cal D}$ is a Wigner rotation matrix. In a similar manner, the
dipole moment ${\vec \mu}$ is determined by its spherical components
such that
\begin{eqnarray}
  -{\vec \mu} \cdot {\vec {\cal F}} &=& - \sum_\iota (-1)^\iota \mu_\iota {\cal
    F}_{-\iota} \nonumber \\ &=& - {\cal F} \sum_{\iota} (-1)^\iota \mu_\iota
  {\cal D}^{1*}_{-\iota0} (\omega_r t,\theta_r, 0) \nonumber \\
  &=& - {\cal F} \sum_\iota \mu_\iota e^{-i \iota \omega_r t}d^1_{-\iota0}(\theta_r).
\end{eqnarray}
This uses the explicit expression for ${\cal D}$ in terms of a
little-$d$ function \cite{bns}.

Just as we treated the internal degrees of freedom in $H_0$ we must
now treat the field interaction.
\begin{eqnarray}
  \langle (\kappa ) j m_{\zeta} | -{\vec \mu} \cdot {\vec {\cal F}} |
  \psi \rangle &=& - {\cal F} \sum_{\kappa^{\prime} j^{\prime}
    m_{\zeta}^{\prime} \iota} (-1)^\iota \langle (\kappa ) j m_{\zeta} | \mu_\iota |
  (\kappa^{\prime}) j^{\prime} m_{\zeta}^{\prime} \rangle \times \nonumber\\
  &&  C_{\kappa^{\prime}, j^{\prime} , m_{\zeta}^{\prime} }
  d^1_{-\iota0}(\theta_r) e^{i \omega_r t (-\iota - m_{\zeta}^{\prime})}.
\end{eqnarray}
Piecing together the different parts, and multiplying through by
$e^{i m_{\zeta} \omega_r t}$, we arrive at the TDSE for the coefficients $C$:
\begin{eqnarray}\label{Schrodinger_for_C}
  i {\dot C}_{\kappa, j, m_{\zeta}} + \omega_r m_{\zeta} C_{\kappa, j,
    m_{\zeta}} &=& \sum_{\kappa^{\prime}, j^{\prime} }
  \langle (\kappa ) j m_{\zeta} | H_0 |(\kappa^{\prime}) j^{\prime} m_{\zeta}\rangle
  C_{\kappa^{\prime}, j^{\prime} , m_{\zeta} } - \nonumber \\
  && {\cal F} \sum_{\kappa^{\prime} j^{\prime} m_{\zeta}^{\prime} \iota}
  (-1)^\iota \langle (\kappa ) j m_{\zeta} | \mu_\iota | (\kappa^{\prime})
  j^{\prime} m_{\zeta}^{\prime} \rangle C_{\kappa^{\prime}, j^{\prime} ,
    m_{\zeta}^{\prime} }\times \nonumber\\  && d^1_{-\iota0}(\theta_r) 
  e^{i \omega_r t (-\iota - m_{\zeta}^{\prime} + m_{\zeta})}.
\end{eqnarray}
Now, we have not specified what field ${\vec {\cal F}}$ is, nor
which structural degrees of freedom are involved in making the
dipole ${\vec \mu}$, and it does not matter.  All that matters is
that ${\vec \mu}$ is a vector, in which case the Wigner-Eckhart
theorem applies \cite{bns}. In the total angular momentum basis, 
we must have
\begin{equation}
  \langle (\nu ) j m_{\zeta} | \mu_\iota | (\nu^{\prime}) j^{\prime}
  m_{\zeta}^{\prime} \rangle \propto \left( \begin{array}{ccc} j & 1 &
    j^{\prime} \\ -m_{\zeta} & \iota & m_{\zeta}^{\prime} \end{array}
  \right),
\end{equation}
where the proportionality constant involves the reduced matrix
element. Then the conservation of angular momentum implies that 
$m_{\zeta}^{\prime}-m_{\zeta} = -\iota$.
However, this immediately removes the time-dependence in the
exponential term in \eqref{Schrodinger_for_C}. In fact, as alluded to
earlier, this statement says that any angular momentum imparted by the rotating
field must be accounted for in the projection $m_\zeta$.

In some cases it will prove more useful to keep track of the individual spin
components $m_{\zeta i}$ separately. For instance, suppose there were two angular
momenta, $m_{\zeta1}$ and $m_{\zeta2}$: we would have two projection terms that 
would each evolve as $e^{i\;m_{\zeta1}\omega_r t}$ and 
$e^{i\;m_{\zeta2}\omega_r t}$. Terms in the Hamiltonian which describe the 
interaction of the two spins are of the form ${\vec s}_1\cdot {\vec s}_2$ for 
which we find scales as
\begin{eqnarray}\nonumber
  &\langle s_1\;m_{\zeta1}\;s_2\;m_{\zeta2}|{\vec s}_1\cdot {\vec s}_2|
  s_1^\prime\;m_{\zeta1}^\prime\;s_2^\prime\;m_{\zeta2}^\prime\rangle \propto&\\
  &\left(\begin{array}{ccc} s_1 & 1 & s_1^\prime\\
    -m_{\zeta1} & p & m_{\zeta1}^\prime\end{array}\right)
    \left(\begin{array}{ccc} s_2 & 1 & s_2^\prime\\
      -m_{\zeta2} & -p & m_{\zeta2}^\prime\end{array}\right)
      e^{i\;(m_{\zeta1}-m_{\zeta1}^\prime)\omega_r\;t}
      e^{i\;(m_{\zeta2}-m_{\zeta2}^\prime)\omega_r\;t}.&
\end{eqnarray}
The proportionality involves a reduced matrix element.
By the conservation of angular momentum we find that
$m_{\zeta1}-m_{\zeta1}^\prime = -(m_{\zeta2}-m_{\zeta2}^\prime)$
and the phase factor is still canceled out. In fact, for any such 
interaction between two spins, the conservation of angular momentum 
forces the time dependence to cancel out. 

With the time-dependence removed, Eq.~\eqref{Schrodinger_for_C}
reduces to the Schr\"{o}dinger equation for a non-rotating field
tilted at an angle $\theta_r$ from the rotation axis, just as in the
spin-$1/2$ example in the prior subsection. This introduces an
additional term on the left of \eqref{Schrodinger_for_C}, which is
moved to the RHS and interpreted as an effective
Hamiltonian. Thus if $H_0$ is presented in the basis 
$| (\kappa )j\;m_j \rangle$ diagonal with respect to the field, 
then the matrix to be diagonalized is
\begin{eqnarray}
  H_{\rm dressed} &=& H_0 - {\vec \mu} \cdot {\vec {\cal F}} -\omega_r m_{\zeta}
  \nonumber \\
  &=& H_0 - {\vec \mu} \cdot {\vec {\cal F}} -\omega_r j_{\zeta},
\end{eqnarray}
where $m_{\zeta}$ is the eigenvalue of the $j_{\zeta}$ operator. As
before, we can rotate this Hamiltonian from the $m_{\zeta}$ basis to
the $m_j$ basis. Since $H_0$ does not depend on the either
$m_{\zeta}$ or $m_j$, it is unaffected by this rotation. In the
frame of the instantaneous field where $m_j$ is the good quantum
number, we can write the dressed Hamiltonian as
\begin{eqnarray}\label{big_result}
  H_{\rm dressed} &=& H_0 - {\vec \mu} \cdot {\vec {\cal F}} -
  {\hat \omega}_r \cdot {\vec j} \nonumber, \\
  &=& H_0 - {\vec \mu} \cdot {\vec {\cal F}} - \omega_r(\cos (\theta_r)j_z-
  \sin(\theta_r)j_x).
\end{eqnarray}
This dressed Hamiltonian is the main result of this article. This Hamiltonian 
has been previously formulated in NMR studies~\cite{Appelt1994}.
In the following sections we will apply it to several elementary 
cases of interest.

\section{Pure Spin-$s$ system}\label{pure-S-section}

As the simplest application of the general method beyond the
spin-$1/2$ particle, we consider in this section a structureless
particle of arbitrary spin $s$, as was considered in the original
formulation of Berry \cite{berry84}.  This spin interacts with a magnetic field
that rotates at an angle $\theta_r$ with respect to the axis of
rotation. Using the result from \eqref{big_result}, this system is
described by the Hamiltonian
\begin{equation}\label{pure-S}
  H_{\rm dressed} = \omega_L s_z - \omega_r(\cos(\theta_r)s_z - \sin(\theta_r)s_x),
\end{equation}
where $\omega_L = g_s \mu_B {\cal B}$ is the $m$-independent Larmor
precession frequency and $g_s$ is the $g$-factor for the spin-$s$.
For this section, we have reverted to the usual
notation $s$ and $m$ for the spin and it's projection onto the
instantaneous field axis.

For this structureless particle, the Hamiltonian (\ref{pure-S}) is
represented by a $(2s+1)$ $\times$ $(2s+1)$ tridiagonal matrix, in
the basis of states $|s\,m \rangle$. This matrix is explicitly given
by
\begin{equation}
  H = \left(\begin{array}{ccccc}
    m\;a & b(s,m) & 0 & 0 & \hdots \\
    b(s,m) & (m-1)\;a & b(s,m-1) & 0 & \vdots \\
    0 & b(s,m-1) & \ddots & \ddots & b(s,-m+1) \\
    \vdots & 0 & b(s,-m+1) & -(m-1)\;a & b(s,-m) \\
    \hdots & 0 & 0 & b(s,-m) & -m\;a. \end{array}\right),
\end{equation}
where $a = \omega_L -\omega_r\cos(\theta_r)$
and $b(s,m)=(1/2)\sqrt{s(s+1)-m(m-1)}\;\omega_r\sin(\theta_r)$.
Appendix A sketches a derivation of the eigenvalues of this matrix,
which are
\begin{equation}\label{pure-S-EV}
  \lambda_m = m\sqrt{\omega_L^2 + \omega_r^2 - 2\omega_r\omega_L\cos(\theta_r)},
\end{equation}
where $m$ takes on the values $-s,\dots,+s$ in integer steps.

The usual Berry phase is obtained in the adiabatic limit where
$\omega_r \ll \omega_L$, in which case
\begin{eqnarray}
  \lambda_{m} \approx m\omega_L - m\omega_r\cos(\theta_r).
\end{eqnarray}
The magnetic field completes one rotation in a time $\tau = 2 \pi /
\omega_r$.  In this time the spin accumulates a dynamical phase
$\varphi_m = m\omega_L \tau$.  Beyond this, it acquires an additional phase
$\gamma_m = \phi_m - \varphi_m$, where $\phi_m = \lambda_m\tau$, 
given to lowest order by
\begin{equation}\label{0oc}
  \gamma_m^{(0)} = - m 2\pi\cos(\theta_r) \Rightarrow 2\pi
  m(1-\cos(\theta_r)).
\end{equation}
In the final step, we use the fact that adding $2\;\pi\; m$ (where $m$
is either integer or half-integer) amounts to adding an unobservable
phase of $\pm1$ to the system. We see that the phase $\gamma_m^{(0)}$ accumulated is
exactly that given by the result of Berry; $m$ times the solid angle
subtended by the rotation. We can also extend this solution to
regimes of non-adiabaticity. The first order correction in
$\omega_r/\omega_L$ is
\begin{equation}\label{1oc}
  \gamma_m^{(1)} =
  2\pi\;m\frac{\omega_r}{2\omega_L}\sin^2(\theta_r),
\end{equation}
which has already been identified elsewhere \cite{berry87,moore91-2}.
Based on our explicit formula, we can extract corrections to any desired 
order, at least for fields undergoing the simple motion in Fig.~\ref{fieldfig}. 
This additional phase can be expanded to any desired 
order in the adiabatic parameter $\omega_r/\omega_L$. For example, 
the second and third-order $\gamma^{(k)}$ corrections are
\begin{eqnarray}
  \gamma_m^{(2)} &=& 
  2\pi m\frac{\omega_r^2}{2\omega_L^2}\cos(\theta_r)\sin^2(\theta_r)\\
  \gamma_m^{(3)} &=& 
  2\pi m\frac{\omega_r^3}{16\omega_L^3}(3+5\cos(2\theta_r))\sin^2(\theta_r)
\end{eqnarray}

Using the general dressed formalism, the limit of fast field rotation can
also be described. The phase $\gamma_m$ (after one field period) can be
approximated in this limit ($\omega_r \gg \omega_L$):
\begin{eqnarray}\label{fastrotation}
  \gamma_m \approx 2\pi m -2\pi m\frac{\omega_L}{\omega_r}\cos(\theta_r),
\end{eqnarray}
and the first term is unobservable. In this case the dominant energy,
as manifested in the phase, is the photon energy due to the time-periodic
field. On top of this, the magnetic field interaction itself makes a small
correction.  This is clearly not the appropriate limit in which to perform 
precision spectroscopy, since small uncertainties in the field rotation 
rate would dominate the observable Larmor frequency.

The structureless spin problem can be solved analytically for arbitrary 
rotation rates since the spin precesses about an \emph{effective} magnetic 
field in the rotating coordinate system~\cite{Rabi1954}.  The effective 
magnetic field makes an angle with respect to the positive $\zeta$-axis, 
$\theta_r^*$, satisfying
\begin{eqnarray}
    \sin(\theta_r^*) &=& 
    \frac{\sin(\theta_r)}{\sqrt{1-2\frac{\omega_r}{\omega_L}\cos(\theta_r) + 
	\left( \frac{\omega_r}{\omega_L} \right)^2}}, \nonumber \\
    \cos(\theta_r^*) &=& \frac{\cos(\theta_r) - 
      \frac{\omega_r}{\omega_L}}{\sqrt{1-2\frac{\omega_r}{\omega_L}\cos(\theta_r) 
	+ \left( \frac{\omega_r}{\omega_L} \right)^2}}.
\end{eqnarray}
This angle smoothly transitions from $\theta_r^* \approx \theta_r$ for 
$\omega_r \ll \omega_L$ to $\theta_r^* \approx \pi$ for $\omega_r \gg \omega_L$.  
Equivalently, the angular deviation, $\Delta\theta_r = \theta_r^* - \theta_r$, 
between the effective magnetic field and the true magnetic field can be 
described through
\begin{eqnarray}
    \sin(\Delta\theta_r) &=& 
    \frac{\frac{\omega_r}{\omega_L}\sin(\theta_r)}{\sqrt{1-2\frac{\omega_r}{\omega_L}
	\cos(\theta_r)+\left( \frac{\omega_r}{\omega_L} \right)^2}},\nonumber \\
    \cos(\Delta\theta_r) &=& 
    \frac{1-\frac{\omega_r}{\omega_L}\cos(\theta_r)}{\sqrt{1-2\frac{\omega_r}{\omega_L}
	\cos(\theta_r)+\left( \frac{\omega_r}{\omega_L} \right)^2}},
\end{eqnarray}
where this deviation smoothly transitions from $\Delta\theta_r^* \approx 0$ 
for $\omega_r \ll \omega_L$ to $\Delta\theta_r^* \approx \pi - \theta_r$ for 
$\omega_r \gg \omega_L$.

The total phase shift, $\phi_m = \varphi_m^\star + \gamma_m^\star$, 
accumulated during one revolution of the magnetic field can be broken up 
into a dynamic contribution, $\varphi_m^\star$, and a geometric 
contribution, $\gamma_m^\star$, where~\cite{Appelt1994,Appelt1995,Wackerle1998}
\begin{eqnarray}
    \varphi_m^\star &=& 
    2 \pi m \frac{\omega_L}{\omega_r} \cos(\Delta\theta_r), \nonumber \\
    \gamma_m^\star  &=& 
    2 \pi m \left( 1-\cos(\theta_r^*) \right).
\end{eqnarray}
$\varphi_m^\star$ and $\gamma_m^\star$ are slightly different than the 
phases defined in Eq.~\eqref{gamma_def}. In this definition, both the 
dynamical phase and the geometric phase acquire non-adiabatic corrections. 
In the limit of very slow rotation $\omega_r$, we have that 
$\varphi_m^\star\rightarrow\varphi_m$. Grouping terms together and 
simplifying provides the total accumulated phase $\phi_m$:
\begin{eqnarray}
    \phi_m &=& 
    2 \pi m \left( 1 + \frac{\omega_L}{\omega_r}\sqrt{1-2\frac{\omega_r}{\omega_L}
      \cos(\theta_r)+\left( \frac{\omega_r}{\omega_L} \right)^2} \right), \nonumber \\
    &=& 2 \pi m \left( 1 + \sqrt{1-2\frac{\omega_L}{\omega_r}\cos(\theta_r) + 
      \left( \frac{\omega_L}{\omega_r} \right)^2} \right).
\end{eqnarray}
The total phase shift arrived at in this geometrical way is the same as the 
dressed state derivation --- $(2\pi/\omega_r)\lambda_m$, where $\lambda_m$ is 
defined in \eqref{pure-S-EV} --- apart from a factor of $2 \pi m$, which is 
unobservable for integer or half-integer values of $m$. This is a semi-classical 
geometric procedure that yields the non-adiabatic corrections to Berry's 
result. 

For the above calculations, we have taken $\vec{\mu} = -g_s \mu_B \vec{S}$ 
and $\omega_r$ to have a \emph{positive} sense about the $\zeta$-axis.  
For the single revolution phase shifts, $\gamma_m^{(i)}$, terms proportional 
to \emph{odd} powers of $\omega_r/\omega_L$ change sign when the g-factor 
changes sign, while terms proportional to \emph{even} powers of 
$\omega_r/\omega_L$ change sign when the sense of rotation changes sign.  
For the expressions defining the angle, $\theta_r^*$, and angular deviation, 
$\Delta\theta_r$, changing the sign of the g-factor (sense of rotation) 
directly changes the sign of $\omega_L$ ($\omega_r$).

\begin{figure}\label{pure-S-fig}
  \begin{center}
    \resizebox{4in}{!}{\includegraphics{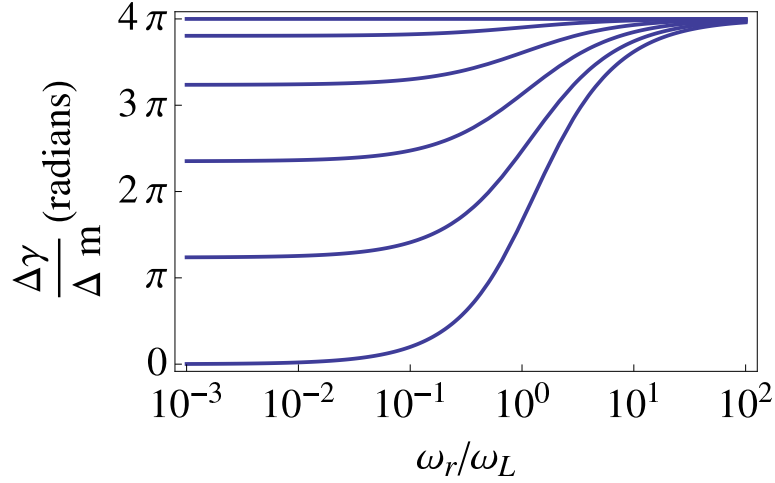}}
  \end{center}
  \caption{The extra phase accumulated due to the rotation of the field. In the limit
  of very fast rotation, $\omega_r\gg\omega_L$, the system accumulates a phase of 
  $4\pi$, which is unobservable. In this limit, the states are best represented by 
  projections onto the axis of rotation. The various lines represent values of 
  $\theta_r$ between $\pi/2$ (bottom line) and $0$ (top line) in steps of $\pi/10$. 
  As can be seen, when $\theta_r$ is zero, there is no measurable phase shift, 
  since there is no enclosed solid angle.}
\end{figure}

A spectroscopic measurement would involve finding the energy difference 
between two states with different values of $m$, with difference $\Delta m$. 
In a Ramsey-type experiment, this measurement seeks to measure the phase 
difference $\omega_L \Delta m\tau = \Delta\varphi$. In a rotating field, 
however, the experiment will produce a measurement of $\Delta\phi = 
\Delta m\sqrt{\omega_L^2+\omega_r^2-2\omega_L\omega_r\cos(\theta_r)}\tau$, 
and thus will introduce an error. This error is given by the difference 
$\Delta \gamma = \gamma_m - \gamma_{m^{\prime}} = \Delta\phi-\Delta\varphi$, 
and is plotted in Fig.~\ref{pure-S-fig} as a function of rotation rate. The 
different curves represent different values of the tilt angle $\theta_r$.  

\section{Structured Spin-$J$ system}

More generally, atoms and molecules are composite objects made of
individual spins, which are moreover coupled together to create 
fine or hyperfine structure.  For example,  alkali atoms couple
the electronic and nuclear spins into a total hyperfine state. 
The resulting angular momentum structure will have a bearing on
the non-adiabatic corrections to the geometric phase accumulated.

As a simple illustration of our formalism, we consider a composite
particle composed of two spin-1/2 objects. This example goes
beyond the structureless particle often envisioned by the usual
Berry theory. The dressed Hamiltonian is given by
\begin{equation}\label{AS-hammy}
  H_{\rm dressed} =
  \omega_{1}j_{1z} + \omega_{2}j_{2z} + \Delta {\vec j}_1 \cdot {\vec j}_2 -
  \omega_r(\cos(\theta_r)J_z - \sin(\theta_r)J_x),
\end{equation}
where $\omega_{i} = g_i \mu_B {\cal B}$ is the Larmor precession frequency
of spin $j_i$; ${\vec J}$ is the vector sum of ${\vec j}_1$ and ${\vec j}_2$;
and $\Delta$ is parameter that governs the splitting between levels $J=0$
and $J=1$.

The Hamiltonian \eqref{AS-hammy} can be represented by a $4\times4$ matrix,
in the basis $\{ |(j_1\,j_2)J\,M_J\rangle \} =$
$\left\{|(\frac{1}{2}\;\frac{1}{2})0\;0\rangle\right.$,
$|(\frac{1}{2}\;\frac{1}{2})1\;1\rangle$, $|(\frac{1}{2}\;\frac{1}{2})1\;0\rangle$,
$\left.|(\frac{1}{2}\;\frac{1}{2})1\;-1\rangle\right\}$
\begin{equation}\label{AS-hammy-matrix}
  H_{\rm dressed} =
  \left(\begin{array}{cccc}
    -\frac{3\Delta}{4} & 0 & \frac{1}{2}(\omega_1-\omega_2) & 0\\
    0 & \frac{\Delta}{4} + \omega_{\rm Z} - \omega_r \cos(\theta_r) &
    \frac{\omega_r}{\sqrt{2}}\sin(\theta_r) & 0 \\
    \frac{1}{2}(\omega_1-\omega_2) & \frac{\omega_r}{\sqrt{2}}\sin(\theta_r) &
    \frac{\Delta}{4} & \frac{\omega_r}{\sqrt{2}}\sin(\theta_r)\\
    0 & 0 & \frac{\omega_r}{\sqrt{2}}\sin(\theta_r) &
    \frac{\Delta}{4}-\omega_{\rm Z}+\omega_r\cos(\theta_r)
    \end{array}\right),
\end{equation}
where $\omega_{\rm Z}=\frac{1}{2}(\omega_1+\omega_2)$, is the average of 
the individual Larmor frequencies. The first item to note is that if the two 
spins have identical Larmor frequencies, $\omega_1 = \omega_2$, then this 
Hamiltonian is equivalent to that of a spin-$0$ particle and a spin-$1$ 
particle that are independent of each other, there is no coupling between 
the two states. Each would then evolve according to the previous section on 
pure spins. This would be the case for the singlet and triplet excited 
states of the helium atom, $(1s2s)^{1,3}S$ state, for example. However, 
should these spins be different from one another (such that 
$\omega_1\ne\omega_2$) then coupling corrections arise.

The ordinary adiabaticity criterion specifies that the rotational frequency 
$\omega_r$ be small as compared to the Larmor precession frequency $\omega_L$, 
which in this example is given by $\omega_{\rm Z}$. However, now it 
becomes also necessary to specify whether the Larmor frequency itself is 
large or small compared to the splitting $\Delta$ between adjacent $J$-levels. 
This is because the Berry phase arises from a correction to the eigenvalues 
of the dressed Hamiltonian relative to the non-rotating Hamiltonian. It is 
therefore worthwhile to cast the non-rotating Hamiltonian in the basis in 
which it is as diagonal as possible. In the following sections we treat the 
two limits separately. Since our emphasis here is on the Berry-phase limit, 
we consider only the limit where $\omega_r \ll \omega_{\rm Z}$, where 
the rotation rate of the field is small compared to the Larmor frequency.
The resulting phase shifts are of course implicit in the theory, however.

\subsection{Weak magnetic field, $\omega_r \ll \omega_{\rm Z} \ll \Delta$}

In the low-field limit, but assuming that each Larmor frequency $\omega_i$ is 
still far larger than the rotational frequency $\omega_r$, we can write down 
expressions for the energy quite simply. Note that in the absence of rotation, 
the leading-order energy shift is the sum of the Larmor frequencies themselves, 
i.e., $(1/2)(\omega_1 + \omega_2)M_J = \omega_{\rm Z}\,M_J$. Leaving this 
correction on the diagonal to break the degeneracy of the $J=1$ level, we now 
treat as perturbations the difference $(1/2)(\omega_1-\omega_2)$ and the rotation 
rate $\omega_r$. 
 
Doing so, the leading-order correction due to rotation of the field 
is given by the diagonal terms in (\ref{AS-hammy-matrix}) that contain 
the rotation rate $\omega_r$. This correction is the usual Berry phase 
found above,
\begin{eqnarray}
  \gamma^{(0)}(|J\;M_J\rangle) = -2 \pi M_J \cos \theta_r,
\end{eqnarray}
and it depends on the atomic state only through the total projection of
angular momentum $M_J$.  Thus the ordinary Berry phase in the limit
of zero rotation rate is still intact, and is independent of the
internal structure.

However, the higher order corrections do depend on this structure. To 
leading order in the rotation frequency $\omega_r$, we find a correction 
to the Berry phase in the 
$|(\frac{1}{2}\;\frac{1}{2})1\;1\rangle$ state:
\begin{equation}
  \gamma^{(1)}(|11\rangle) = 2\pi\frac{\omega_r}{2\omega_{\rm Z}}
  \sin^2(\theta_r)\left(1+\frac{(\omega_1-\omega_2)^2}{4\omega_{\rm Z}
    (\Delta+\omega_{\rm Z})}
  \right).
\end{equation}
The first term is the usual first order correction for a structureless 
particle (cf. \eqref{1oc}), 
with the replacement of $\omega_L$ by $\omega_{\rm Z}$. This should 
be expected since the energy splitting between the two states is 
given by $\omega_{\rm Z}$, and thus is what must be overcome by 
the rotating field that couples together the differing projections. The second 
term in the parentheses depends on how strongly the rotating field couples states 
with differing total $J$, as manifested by $\omega_1 - \omega_2$. This 
new correction arises from 4$^{\rm th}$-order mixing in perturbation theory, 
it is nevertheless linear in the adiabaticity parameter 
$\omega_r/\omega_{\rm Z}$. A similar expression is found for the 
$|1-1\rangle$ state,
\begin{equation}
  \gamma^{(1)}(|1-1\rangle) = -2\pi\frac{\omega_r}{2\omega_{\rm Z}}
  \sin^2(\theta_r)\left(1-\frac{(\omega_1-\omega_2)^2}{4\omega_{\rm Z}
    (\Delta-\omega_{\rm Z})}
  \right).
\end{equation}
As is expected from the pure spin case, this state picks up an overall negative 
sign. However, due to slight changes introduced by the structure,
we find a slightly different correction to the second term in parentheses. In fact, 
we can write down an expression that encapsulates the first order 
(in $\omega_r/\omega_{\rm Z}$) correction as
\begin{equation}
  \gamma^{(1)}(|J\;M_J\rangle) = 2\pi M_J \frac{\omega_r}{2\omega_{\rm Z}}
  \sin^2(\theta_r)\left(1+M_J\frac{(\omega_1-\omega_2)^2}{4\omega_{\rm Z}
    (\Delta+M_J\omega_{\rm Z})}
  \right).
\end{equation}
While the first term is exactly of the form in \eqref{1oc}, the second term 
describes how the distant $|J=0,M_J=0\rangle$ state affects 
the accumulated first order phase $\gamma_{M_J}^{(1)}$; namely that the 
quadratic Zeeman shift in the two $M_J=0$ levels distorts the system 
such that the $|J=1,M_J=1(-1)\rangle$ state is affected more (less) by the 
$|J=1,M_J=0\rangle$ state. 

The two states with $M_J=0$ do not acquire a geometric phase at lowest 
order in $\omega_r$, which is appropriate. In this case the leading order 
perturbation to the dressed Hamiltonian is 
$E_{\rm Z}^{(\pm)} = \pm [(1/2)(\omega_1 - \omega_2)]^2/\Delta$,
which denotes the quadratic Zeeman shift already present in the non-rotating 
system, and which does not contribute to the Berry phase $\gamma$. The 
quadratic shift after a period $\tau=2\pi/\omega_r$ is the dynamical phase 
the $M_J=0$ would nominally acquire. To the first order in $\omega_r$ in which 
there is a correction to the $|J=0,M_J=0\rangle$ state arises in 
4$^{\rm th}$-order perturbation theory. It is given by 
\begin{equation}\label{1oc00}
  \gamma^{(1)} (|0\,0\rangle) = 2\pi\frac{\omega_r}{2\Delta}\sin^2(\theta_r) 
  \frac{2E_{\rm Z}^{(-)}}{\Delta\left(1-\left(\frac{\omega_{\rm Z}}{\Delta}
    \right)^2\right)},
\end{equation}
where as always the superscript ``1'' denotes a correction linear in 
$\omega_r$. Here we find a term that appears similar to the $M_J=1$ 
states, with the exception that it occurs in an $M_J=0$ state. We have 
introduced a new energy scale into the problem by adding $\Delta$ and this 
allows the $|0\,0\rangle$ state to acquire a first order Berry phase. However, 
the strength of this phase is reduced by a term proportional to the ratio 
of the quadratic Zeeman shift in the lower level to the spin-spin energy 
splitting. Given our assumptions, this term --- while linear in $\omega_r$ --- 
is a product of multiple small parameters, and is generally smaller 
than $\gamma^{(0)}$ for $M_J=\pm1$ states.

For the case of the $|J=1,M_J=0\rangle$ state, there is also a $4^{\rm th}$-order 
correction, but it takes a very different form. We find, after much algebra, 
\begin{equation}\label{1oc10}
  \gamma^{(1)} (|1\,0\rangle) = -2\pi\frac{\omega_r}{2\omega_{\rm Z}}\sin^2(\theta_r)
  \frac{2E_{\rm Z}^{(+)}}{\omega_{\rm Z}}.
\end{equation}
This is very different from the $|0\,0\rangle$ state correction in \eqref{1oc00}. 
The important energy scale is the linear Zeeman shift $\omega_{\rm Z}$. 
Eqs.\,\eqref{1oc00} and \eqref{1oc10} carry an important insight; the energy scale 
responsible for higher-order Berry phases is different for the two $M_J=0$ states. 
The dominant scale in the $|0\,0\rangle$ is the spin-spin splitting $\Delta$. In 
the $|1\,0\rangle$, the dominant energy scale is the linear Zeeman shift. To 
first order in $\omega_r$, there is a correction to the $M_J=0$ states 
that, while similar to the shift in the $|M_J|=1$ states, is reduced in magnitude. 
This reduction is due to the structure, the structure that provides a quadratic 
Zeeman shift in the $M_J=0$ states. For the lower (upper) level, the correction 
depends on the relative strength of the quadratic Zeeman shift to the 
spin-spin splitting (linear Zeeman shift). In the regime considered, both 
of these contributions are very small. The same ideas apply to the $F=0$ and 
$F=1$ hyperfine states of Hydrogen, where the magnetic field is coupling states 
of the same parity. Briefly, $\gamma^{(1)}$ is influenced by ``nearby'' $M_J=\pm1$ 
states for the $|1\,0\rangle$ level, and comparitively less influenced by the 
``far away'' $M_J=\pm1$ levels in the $|0\,0\rangle$ state.

It is instructive to examine these results for different cases of individual 
Larmor frequencies.  In the case where both particles experience the same
Larmor frequency in a field, $\omega_1 = \omega_2$, then these first-order
corrections reduce to the usual first-order corrections for a structureless
spin-1 particle, as in Eqn.\,\eqref{1oc}, and the additional $M_J=0$ pieces are
zero as well.  In another limit where one Larmor frequency dominates the other, 
say $\omega_1 \gg \omega_2$, then $\gamma^{(1)}$ for the $|M_J|=1$ states reduce 
to the first order correction of the dominant spin alone, reflecting the 
fact that the weaker spin is coupled to the stronger one and gets dragged 
along for the ride. This happens, for example, in the $F=1$ hyperfine ground 
state of the hydrogen atom, where the nuclear g-factor is far smaller than the 
electron g-factor.

\subsection{Strong magnetic field, $\omega_1, \omega_2 \gg \Delta \gg \omega_r$}

In the other limit, where the magnetic field is large compared to the splitting
between adjacent $J$-levels, it is more useful to construct the dressed
Hamiltonian in an alternative basis.  Namely, the non-rotating
Hamiltonian is more nearly diagonal in the independent-spin basis 
$|j_1\,m_1\,j_2\,m_2\rangle$, where the four Zeeman energies $E_{m1,m2}$ 
are given simply by $m_1 \omega_1 + m_2 \omega_2$:
\begin{eqnarray}\nonumber
  E_{\frac{1}{2},\frac{1}{2}} &=& \frac{1}{2}(\omega_1+\omega_2) \\
  \nonumber
  E_{\frac{1}{2},-\frac{1}{2}} &=& \frac{1}{2}(\omega_1-\omega_2) \\
  \nonumber
  E_{-\frac{1}{2},-\frac{1}{2}} &=& -\frac{1}{2}(\omega_1+\omega_2) \\
  \label{highB-e-comp}
  E_{-\frac{1}{2},\frac{1}{2}} &=& -\frac{1}{2}(\omega_1-\omega_2) ,
\end{eqnarray}
as appropriate to this Paschen-Back limit of the Zeeman effect. This is an
example of the aside in Sec.\;II where we make the {\it ansatz} $(\alpha,\;\beta)$
$e^{-i\;m_1\omega_r t}e^{-i\;m_2\omega_r t}$. The remaining Hamiltonian,
which includes the rotation of the field and the spin-spin interaction,
is recast as follows
\begin{equation}\label{V-recast}
  H_{\rm dressed} = \left(\begin{array}{cccc}
    E_{\frac{1}{2},\frac{1}{2}} + \frac{\Delta}{4}-\omega_r\cos(\theta_r) &
    \frac{\omega_r}{2}\sin(\theta_r) & 0 & -\frac{\omega_r}{2}\sin(\theta_r)\\
    \frac{\omega_r}{2}\sin(\theta_r) & E_{\frac{1}{2},-\frac{1}{2}} - 
    \frac{\Delta}{4} & \frac{\omega_r}{2}\sin(\theta_r) & -\frac{\Delta}{2}\\
    0 & \frac{\omega_r}{2}\sin(\theta_r) & E_{-\frac{1}{2},-\frac{1}{2}} +
    \frac{\Delta}{4} + \omega_r\cos(\theta_r) & -\frac{\omega_r}{2}\sin(\theta_r)\\
    -\frac{\omega_r}{2}\sin(\theta_r) & -\frac{\Delta}{2} &
    -\frac{\omega_r}{2}\sin(\theta_r) & E_{-\frac{1}{2},\frac{1}{2}} - 
    \frac{\Delta}{4} \end{array}\right)
\end{equation}
Once again, we can immediately read the Berry-phase contribution from the
diagonal components, as
\begin{eqnarray}\nonumber
  \gamma^{(0)}(|j_1\;m_1\;j_2\;m_2\rangle)&=& -2\pi(m_1+m_2)\cos(\theta_r)
\nonumber \\
&=& -2 \pi M_J \cos (\theta_r).
\end{eqnarray}
The phase accumulates due to the individual spins separately, as expected when the
spins interact weakly with each other compared to their interaction with the field.

This independent accumulation of phase leads to a different interpretation of the
$M_J=0$ states: in the limit of small magnetic field compared to the spin-spin 
energy splitting, we had attributed this to an $M_J=0$ projection while here, we 
can attribute this to $m_1 = \pm\frac{1}{2}$ accumulating 
$\pm(\omega_r/2)\cos(\theta_r)$ extra energy and the $m_2 = \mp\frac{1}{2}$ 
accumulating $\mp(\omega_r/2)\cos(\theta_r)$. To cement this idea even further, 
we find there are two independent first-order contributions to the first-order 
non-adiabatic correction $\gamma^{(1)}$, computed first neglecting $\Delta$:
\begin{eqnarray}
  \gamma^{(1)}(|j_1\;m_1\;j_2\;m_2\rangle) &=& 
  2\pi\;m_1\frac{\omega_r}{2\omega_{1}}\sin^2(\theta_r)
  + 2\pi\;m_2\frac{\omega_r}{2\omega_{2}}\sin^2(\theta_r)\\ 
  &=& \gamma_1^{(1)}(|j_1\;m_1\rangle)+\gamma_2^{(1)}(|j_2\;m_2\rangle).
\end{eqnarray}
This is exactly the contribution one would expect from two {\it independent} spins
following a rotating field. When the spins are anti-aligned, or $m_1=-m_2$, we 
find no correction at this order. It is worth noting that this perturbative 
expansion breaks down if $\omega_r \sim \omega_i$. Thus, should the rotation rate
be fast with respect to one of the Larmor frequencies, but not the other, then
the measured phase difference cannot be treated perturbatively in this regime.

The explicit effect of internal structure, manifested in the splitting $\Delta$,
appears as a next-order correction:
\begin{eqnarray}
  \gamma_1^{(1)} &\rightarrow& 2\pi\;m_1\frac{\omega_r}{2\omega_1}\sin^2(\theta_r)
  \left(1+\left(\frac{\Delta}{2\omega_2}\right)^2\right)\\
  \gamma_2^{(1)} &\rightarrow& 2\pi\;m_2\frac{\omega_r}{2\omega_2}\sin^2(\theta_r)
  \left(1+\left(\frac{\Delta}{2\omega_1}\right)^2\right).
\end{eqnarray}
Of course, there are many routes by which $4^{\rm th}$-order perturbation 
theory can affect this state. We only give one route which produces a phase 
shift proportional to $\omega_r$ after one period of oscillation. There is a 
structure correction for each non-adiabatic spin that depends on the relative 
strength of the spin-spin splitting to the other Larmor frequency. Thus, we 
require that the spin-spin splitting be small compared to each of the Larmor 
frequencies in order to make this expansion. Again, this is a more restrictive 
condition on adiabaticity than is usually employed for two independent
spins.

\section{Polar Molecules in a rotating electric field}

Molecules bring yet another degree of freedom to the picture,
namely end-over-end rotation with eigenstates $|N\;M_N\rangle$.
In addition, if the molecule is polar, it has an electric
dipole moment that can be acted upon by a rotating electric field. In this
section we will consider only diatomic molecules, and only one of a fairly
simple structure, to illustrate how our formalism applies to them. The 
lowest-order Berry phase was worked out recently in this system 
\cite{DeMille09}, but the higher-order corrections are 
implicit there as well.

For the sake of illustration we choose the simplest of diatomic molecules, 
a $^1\Sigma$ molecule with no hyperfine structure. In a rotating electric 
field this system is described by a Hamiltonian of the form
\begin{equation}\label{1sigmaMol}
  H_{\rm dressed} = B{\vec N}^2 - {\vec \mu}_m\cdot{\vec{\cal E}} -
  \omega_r(\cos(\theta_r)N_z - \sin(\theta_r)N_x),
\end{equation}
where ${\vec N}$ is the end-over-end rotational angular momentum of the 
molecule, $\mu_m$ is the electric dipole moment of the molecule and ${\cal E}$ 
is the electric field strength. Since we are in the frame of the
electric field and the electric dipole moment points along the molecular 
axis, we do not have any couplings of $M_N$ or $\Lambda$, where $\Lambda$ 
is the projection of total angular momentum onto the internuclear axis. 
For $\Sigma$-molecules, this means there are no couplings to excited 
electronic states by the applied electric field at the low fields we 
consider.

For simplicity, we consider here the coupling only between the $N=0$ and $N=1$
rotational levels of the molecule, assuming weak coupling of rotational
states due to the electric field, i.e., $\mu_m {\cal E} \ll B$.
The formalism can of course be extended to arbitrarily large $N$ 
values as needed.  It is nice to note that this formalism has an atomic analog: 
the ${}^1S_0$ and ${}^1P_1$ states of noble gas and alkaline-earth atoms have 
opposite-parity and are coupled by the Stark interaction. This approach gives 
the corrections for states of opposite parity coupled by the Stark interaction. 
The dressed Hamiltonian reads, in the basis 
$\{ |N M_N \rangle \} = \{|00 \rangle, |1-1 \rangle, | 10 \rangle,|1+1 \rangle \}$
\begin{eqnarray}
  H_{\rm dressed} = \left( \begin{array}{cccc}
    0 & 0 & -\frac{1}{\sqrt{3}} \mu_m {\cal E} & 0 \\
    0 & 2B-\omega_r \cos( \theta_r) & \frac{1}{\sqrt{2}} \omega_r \sin (\theta_r)
    & 0 \\ -\frac{1}{\sqrt{3}} \mu_m {\cal E} & \frac{1}{\sqrt{2}} \omega_r 
    \sin (\theta_r) & 2B & \frac{1}{\sqrt{2}} \omega_r \sin (\theta_r)  \\
    0 & 0 & \frac{1}{\sqrt{2}} \omega_r \sin (\theta_r) & 2B
    + \omega_r \cos (\theta_r) \\
  \end{array} \right)
\end{eqnarray}
Note the the electric Hamiltonian is off-diagonal in the basis of parity 
eigen-states. In the absence of the perturbation $\omega_r$, this Hamiltonian 
appears to have a complete degeneracy among the three states with $N=1$.  In the
magnetic field case above, this degeneracy was broken by the linear
Zeeman effect acting on the diagonal matrix elements.  To achieve the
same feat here, we must account for the off-diagonal mixing due to
electric field.  Note the similarity of this procedure to that of 
Vutha and DeMille~\cite{DeMille09}.

We first diagonalize the $M_N=0$ subspace, using the mixing angle $\delta$
defined by
\begin{eqnarray}
  \tan(\delta) = -\frac{\mu_m {\cal E}}{\sqrt{3}\,B} = -x,
\end{eqnarray}
with the usual eigenvectors $(\cos(\delta/2),\sin(\delta/2))$ and 
$(-\sin(\delta/2),\cos(\delta/2))$. The explicit values in terms 
of the parameter $x$ are 
\begin{eqnarray}
  \cos\left(\frac{\delta}{2}\right) 
  &=& \sqrt{\frac{\sqrt{1+x^2}+1}{2\sqrt{1+x^2}}}\\
  \sin\left(\frac{\delta}{2}\right) 
  &=& \sqrt{\frac{\sqrt{1+x^2}-1}{2\sqrt{1+x^2}}}.
\end{eqnarray}
In terms of this mixing angle the transformed Hamiltonian, with 
electric-field-dependent terms on the diagonal only, reads
\begin{eqnarray}
  H_{\rm dressed} = \left( \begin{array}{cccc}
    B(1 - \sqrt{1 + x^2}) & - \frac{\omega_r}{\sqrt{2}}\sin(\frac{\delta}{2}) 
    \sin(\theta_r) & 0 & - \frac{\omega_r}{\sqrt{2}}\sin(\frac{\delta}{2}) 
    \sin(\theta_r) \\
    - \frac{\omega_r}{\sqrt{2}} \sin(\frac{\delta}{2}) \sin(\theta_r)& 
    2B - \omega_r \cos(\theta_r) & \frac{\omega_r}{\sqrt{2}} \cos(\frac{\delta}{2}) 
    \sin(\theta_r) & 0 \\ 0 & \frac{\omega_r}{\sqrt{2}} \cos(\frac{\delta}{2}) 
    \sin(\theta_r) & B(1 + \sqrt{1 + x^2}) & \frac{\omega_r}{\sqrt{2}} 
    \cos(\frac{\delta}{2}) \sin(\theta_r)   \\
    - \frac{\omega_r}{\sqrt{2}}\sin(\frac{\delta}{2}) \sin(\theta_r)  & 0 
    & \frac{\omega_r}{\sqrt{2}}\cos(\frac{\delta}{2} \sin(\theta_r)   & 
    2B + \omega_r \cos(\theta_r) \\
  \end{array} \right).
\end{eqnarray}
In the limit that $x\ll1$ we see that the diagonal terms for the two $M_N=0$ 
states are merely the quadratic Stark shift 
$E_{\rm S}^{(\pm)} = \pm(\mu_m {\mathcal E})^2/6B$. 
From here we can read off the ordinary Berry 
phase from the diagonal perturbations linear in $\omega_r$, yielding the usual
\begin{eqnarray}
  \gamma^{(0)}(|{\tilde N}\,M_N \rangle) = -2 \pi M_N \cos(\theta_r),
\end{eqnarray}
where by ${\tilde N}$ is meant the appropriate eigenstate of the field-mixed 
$M_N=0$ states~\cite{DeMille09}. This diagonalization removed the degeneracy 
of the $M_N=0$ level with the $M_N=\pm1$ levels of the $N=1$ subspace. However, 
the degeneracy among the $M_N=\pm1$ states still exists.

Having quasi-broken the degeneracy in the $N=1$ levels, we can now evaluate 
the first-order adiabatic correction term using standard second-order 
perturbation theory. It is evident that both states $|N,M_N=\pm 1 \rangle$ 
experience the same additional phase at this order (due to their degeneracy in 
the non-perturbing Hamiltonian), given  by 
\begin{eqnarray}\label{gamma_a_efield}
  \gamma^{(1)}(|1,\,\pm1) = 
  -\pi \left(\frac{6\,B} {(\mu_m {\cal E})^2}\right)\omega_r\sin^2( \theta_r),
\end{eqnarray}
where an expansion in the small parameter $x$ has been applied. Requiring 
this to be a small correction identifies the adiabaticity criterion for
this situation.  If this case were analogous to the magnetic field case, we would
only be concerned about the magnitude of $\omega_r$ with respect to 
$\mu_m {\cal E}$, which is the stand-in for the Larmor frequency in this case. 
However, Eqn.~\eqref{gamma_a_efield} suggests a slightly different criterion, 
namely $\omega_r B \ll (\mu_m{\cal E})^2$ must hold in order to recover the
simple leading-order Berry phase. To understand the origin of this 
criterion we look at the term in the large parentheses in \eqref{gamma_a_efield}.
It is the inverse of the Stark energy in the absence of field rotation for the 
upper level. We can rewrite \eqref{gamma_a_efield} as 
\begin{equation}
  \gamma^{(1)}(|N\,M_N=\pm1\rangle) = -2\pi \frac{\omega_r}{2\,E_{\rm S}}
  \sin^2(\theta_r),
\end{equation}
and we recover a form reminiscent of the pure spin case (cf. \eqref{1oc}), 
where the Larmor frequency, $\omega_L$, is replaced by $E_{\rm S}$, the 
quadratic Stark shift. In order to be an adiabatic correction, it is 
immediately evident why $\omega_r B \ll (\mu_m{\cal E})^2$ must hold; 
the rotation rate must be small compared to the energy splitting in that 
level. In this case, the splitting is quadratic in electric field and 
therefore a secondary energy scale --- the rotational level splitting or 
internal structure --- must come into play. 

By similar reasoning, we can arrive at the $\gamma^{(1)}$ corrections 
for the $M_N=0$ states as well. That of the lower level is given by
\begin{equation}
  \gamma^{(1)}(|{\tilde N}\sim0\,,0\rangle) = 
  -2\pi\frac{\omega_r}{2B}\sin^2(\theta_r)
  \frac{E_{\rm S}}{2\,B},
\end{equation}
where we find the requirement that $\omega_r\,E_{\rm S}\ll B^2$ must hold. 
In this case, our assumptions clearly support this adiabatic criterion  since 
we are in the regime $\omega_r\ll \mu_m {\cal E}\ll B$. This correction is 
linear in $\omega_r$, but suppressed by the ratio of the Stark energy to the 
rotational constant of the molecule. Physically, this is because this state is 
far removed from the ``degeneracy'' in the $N=1$ levels. It is the electric 
analogue of the weak magnetic field limit of the spin-spin interaction. The 
correction for the upper level is given by
\begin{equation}
  \gamma^{(1)}(|{\tilde N}\sim1\,,0\rangle) = 4\pi \frac{\omega_r}{2\,E_{\rm S}}
  \sin^2(\theta_r).
\end{equation}
This correction is in the opposite direction to and twice that of the 
$|N\,M_N=\pm1\rangle$ states because the $|{\tilde N}\sim1\,,0\rangle$ state 
is influenced by the two states, $M_N=\pm1$, that are below it in energy. This 
is in contrast to the coupled spins in a magnetic field case, where each $M_N=\pm1$ 
contributed equally in magnitude but opposite in sign. This is due to the lack 
of any linear Stark shift in the $M_N=\pm1$ levels.

It is evident that polar molecules in a rotating electric field are 
quite similar to magnetic dipoles in rotating magnetic fields. There is 
an energy splitting in comparison with which the rotation of the field must be 
small to ensure adiabaticity. If there is a shift in energy that is linear with the 
applied field, then the rotation rate must be small compared to this energy. 
However, if the energy scales quadratically with the applied field, the rate 
of rotation must be small in comparison to the energy shift in the field. Thus, 
the internal structure is quite important in regimes of quadratic field shifts 
and introduces different adiabaticity requirements on $\omega_r$ in terms 
of the applied field and internal structure.

\section{Conclusions}

The dressed-state formalism is a natural way to treat quantum mechanical
objects subject to time-periodic driving such as we have considered here.
It allows for the inclusion of arbitrary internal structure of the
object, and still reveals the exact dressed eigen-energies at arbitrary
rotation rates.   In the limit of slow rotations, it also reduces,
as it must, to the usual geometric Berry phase.  Because it includes
the structure of the atom or molecule considered, however, it is also
able to shed light on the influence of this structure on non-adiabatic
corrections to the geometric phase.  It is therefore expected to be a
powerful tool to be used when analyzing high-precision spectroscopic
data in the presence of periodic driving 
\cite{Pendlebury04_PRA,Lamoreaux05_PRA}.

The treatment herein has considered only the simplest case of a magnetic
or electric field whose direction precesses uniformly about a given
axis.  It is to be expected, however, that this treatment is yet more
general, and that dressed states for arbitrary periodic driving Hamiltonians 
could be constructed, at least numerically.  It could, for example, be combined
with the Floquet analysis that has been used previously for unstructured
particles \cite{moore90,moore91,moore91-2,stutz}.

\begin{acknowledgments}
  We wish to thank R. Stutz, L. Sinclair, N. Shafer-Ray, and D. DeMille 
for useful discussions.  We acknowledge the financial support of the NSF. 
\end{acknowledgments}

\appendix
\section{}

In Sec.\;\ref{pure-S-section}, we had a Hamiltonian of the form
\begin{equation}
  H = \omega_L S_z - \omega_r(\cos(\theta_r)S_z - \sin(\theta_r)S_x),
\end{equation}
which looks like
\begin{equation}
  H = \left(\begin{array}{ccccc}
    a_1 & c_1 & 0 & 0 & \hdots \\
    c_1 & a_2 & c_2 & 0 & \vdots \\
    0 & c_2 & \ddots & \ddots & c_{n-2} \\
    \vdots & 0 & c_{n-2} & a_{n-1} & c_{n-1} \\
    \hdots & 0 & 0 & c_{n-1} & a_n. \end{array}\right),
\end{equation}
In our system, not only is this matrix symmetric, it contains the following added
symmetry: $a_1 = -a_n$, $a_2 = -a_{n-1}$, etc. Also, the coupling coefficients $c_i$
follow a similar property: $c_1 = c_{n-1}$, etc. These properties are key to
simplifying the eigenvalues of the tridiagonal matrix in our case.

Eigenvalues of an $n\times n$ tridiagonal matrix are given by the roots of the
polynomial $p_n$, defined recursively by
\begin{eqnarray}\nonumber
  p_0(\lambda) &=& 1, \\ \nonumber
  p_1(\lambda) &=& (a_1 - \lambda), \\ \nonumber
  p_2(\lambda) &=& (a_2 - \lambda)p_1(\lambda) - c_1^2 p_0(\lambda) \\ \nonumber
  &\vdots& \\ \label{characteristic}
  p_n(\lambda) &=&
  (a_n - \lambda)p_{n-1}(\lambda) - c_{n-1}^2 p_{n-2}(\lambda).
\end{eqnarray}
In our problem, the constants are defined as
\begin{equation}
  a_m = m(\omega_L - \cos(\theta_r)),
\end{equation}
\begin{equation}
  c_m \sim \left(\begin{array}{ccc}
    S & 1 & S\\
    -m & q & m-q
  \end{array}\right).
\end{equation}
Thus, the symmetry pops right out.

A simple example is the case of $S=1/2$. Here we find that the characteristic
polynomial is
\begin{equation}\label{char2}
  p_2(\lambda)= (\lambda-\frac{1}{2}a)(\lambda+\frac{1}{2}a)-
  \left(\frac{1}{2}b\right)^2  = (\lambda^2-\frac{1}{4}(a^2+b^2))
\end{equation}
where $a = (\omega_L - \omega_r \cos (\theta_r))$ and
$b = \omega_r\sin(\theta_r)$. For the case of $S=1$ we find a
similar equation (after simplification)
\begin{eqnarray}
  p_3(\lambda) &=& \lambda((\lambda-a)(\lambda+a)-2\left(\frac{1}{\sqrt{2}}b\right)^2)
  \nonumber \\ \label{char3}
  &=& \lambda(\lambda^2-(a^2+b^2))
\end{eqnarray}
For integer values there is always a diagonal element that is 0. As is evident, this
has the same form as \eqref{char2} with the added piece of $\lambda$ multiplying
everything yielding an eigenvalue of $0$. In addition, Eq.\;\eqref{char3} is
scaled by a factor of 4 from from Eq.\;\eqref{char2}, thus making the the
eigenvalues a factor of 2 larger. This is because the value of $m$ in \eqref{char3}
is twice as large as the value of $m$ in \eqref{char2}.

We can generalize the characteristic polynomial to a very simple expression
due to the added symmetries. It is given by
\begin{equation}\label{charSim}
  p_{2S+1} = \prod_{m=(m_{\rm min}\ge0)}^{m_{\rm max}}
  (\lambda^{2(1-\delta_{m_{\rm min},0})}-m^2(a^2+b^2)).
\end{equation}
The Kronecker $\delta$-function in \eqref{charSim} is to insure that in
the event $m_{\rm min}=0$ there is only one eigenvalue $\lambda = 0$.


\begin{thebibliography}{10}

\bibitem{berry84}
M.~V. Berry, F.R.S.
%\newblock Quantal phase factors accompanying adiabatic changes.
\newblock {\em Proc.~R.~Soc.~Lond.~A}, 382:45--57, 1984.

\bibitem{simon83}
B.~Simon.
%\newblock Holonomy, the quantum adiabatic theorem, and berry's phase.
\newblock {\em Phys.~Rev.~Lett.}, 51(24):2167--2170, 1983.

\bibitem{anandan87}
J.~Anandan and L.~Stodolsky.
%\newblock Some geometrical considerations of berry's phase.
\newblock {\em Phys.~Rev.~D}, 35(8):2597--2600, 1987.

\bibitem{aharonov87}
Y.~Aharonov and J.~Anandan.
%\newblock Phase change during a cyclic quantum evolution.
\newblock {\em Phys.~Rev.~Lett.}, 58(16):1593--1596, 1987.

\bibitem{page87}
D.~N. Page.
%\newblock Geometrical description of berry's phase.
\newblock {\em Phys.~Rev.~A}, 36(7):3479--3481, 1987.

\bibitem{garrison}
J.~C. Garrison and E.~M. Wright.
%\newblock Complex geometrical phases for dissipative systems.
\newblock {\em Phys.~Lett.~A}, 128(3,4):177--181, 1988.

\bibitem{robbins94}
J.~M. Robbins and M.~V. Berry.
%\newblock A geometric phase for $m=0$ spins.
\newblock {\em J. Phys. A: Math. Gen.}, 27:L345--L348, 1994.

\bibitem{berry87}
M.~V. Berry, F.R.S.
%\newblock Quantum phase corrections from adiabatic iteration.
\newblock {\em Proc.~R.~Soc.~Lond.~A}, 414:31--46, 1987.

\bibitem{berry90}
M.~V. Berry, F.R.S.
%\newblock Histories of adiatbatic quantum transitions.
\newblock {\em Proc.~R.~Soc.~Lond.~A}, 469:61--72, 1990.

\bibitem{berry90-2}
M.~V. Berry, F.R.S.
%\newblock Geomtric amplitude factors in adiabatic quantum transitions.
\newblock {\em Proc.~R.~Soc.~Lond.~A}, 430:405--411, 1990.

\bibitem{cui92}
Shi-Min Cui.
%\newblock Nonadiabatic berry phase in rotating systems.
\newblock {\em Phys.~Rev.~A}, 45(7):5255--5237, 1992.

\bibitem{hannay98}
J.~H. Hannay.
%\newblock The berry phase for spin in the majorana representation.
\newblock {\em J. Phys. A: Math. Gen.}, 31:L53--L59, 1998.

\bibitem{moore90-2}
D.~J. Moore.
%\newblock Berry phases and hamiltonian time dependence.
\newblock {\em J. Phys. A: Math. Gen.}, 23:5523--5534, 1990.

\bibitem{moore91-2}
D.~J. Moore.
%\newblock The calculation of nonadiabatic berry phases.
\newblock {\em Physics Reports}, 210(1):1--43, 1991.

\bibitem{moore90}
D.~J. Moore.
%\newblock Floquet theory and the nonadiabatic berry phase.
\newblock {\em J. Phys. A: Math. Gen.}, 23:L665--L668, 1990.

\bibitem{moore91}
D.~J. Moore and G.~E. Stedman.
%\newblock Adiabatic and nonadiabatic berry phase for two-level atoms.
\newblock {\em Phys.~Rev.~A}, 45(1):513--519, 1991.

\bibitem{horsley07}
S.~A.~R. Horsley and M.~Babiker.
%\newblock Topological phases for composite particles with dynamic properties.
\newblock {\em Phys.~Rev.~Lett.}, 99:090401, 2007.

\bibitem{wilczek84}
F.~Wilczek and A.~Zee.
%\newblock Appearance of gauge strucure in simple dynamical systems.
\newblock {\em Phys.~Rev.~Lett.}, 52(24):2111--2114, 1984.

\bibitem{yaffe87}
J.~E. Avron, R.~Seiler, and L.~G. Yaffe.
%\newblock Adiabatic theorems and applications to the quantum hall effect.
\newblock {\em Commun. Math. Phys.}, 110:33--49, 1987.

\bibitem{wang99}
Z.-C. Wang and B.-Z. Li.
%\newblock Geomtric phase in relativistic quantum theory.
\newblock {\em Phys.~Rev.~A}, 60(6):4313--4317, 1999.

\bibitem{hoodbhoy88}
P.~Hoodbhoy.
%\newblock Berry's phase for atomic levels.
\newblock {\em Phys.~Rev.~A}, 38(7):3766--3768, 1988.

\bibitem{DeMille09}
A.~Vutha and D.~DeMille.
%\newblock Geometric phases without geometry.
\newblock {\em ArXiV:0907.5116}, pages 1--8, 2009.

\bibitem{cohen}
C.~Cohen-Tannoudji, J.~Dupont-Roc, and G.~Grynberg.
\newblock {\em Atom-Photon Interactions}.
\newblock John Wiley \& Sons, Inc., New York, 1992.

\bibitem{qing09}
Q.~Ai, W.~Huo, G.~L. Long, and C.~P. Sun.
%\newblock Non-adiabatic fluctuation in measured geometric phase.
\newblock {\em ArXiV:0903.5381v2}, pages 1--4, 2009.

\bibitem{bns}
D.~M. Brink and G.~R. Satchler.
\newblock {\em Angular Momentum, 3$^{\rm rd}$ Ed.}
\newblock Clarendon Press, Oxford, 1993.

\bibitem{Appelt1994}
S.~Appelt, G.~W\"ackerle, and M.~Mehring.
%\newblock Deviation from berry's adiabatic geometric phase in a $^{131}${X}e
%  nuclear gyroscope.
\newblock {\em Phys.~Rev.~Lett.}, 72(25):3921--3924, 1994.

\bibitem{Rabi1954}
I.~I. Rabi, N.~F. Ramsey, and J.~Schwinger.
%\newblock Use of rotating coordinates in magnetic resonance problems.
\newblock {\em Rev. Mod. Phys.}, 26(2):167--171, 1954.

\bibitem{Appelt1995}
S.~Appelt, G.~W\"ackerle, and M.~Mehring.
%\newblock A magnetic resonance study of non-adiabatic evolution of spin quantum
%  states.
\newblock {\em Z. Phys. D}, 34:75--85, 1995.

\bibitem{Wackerle1998}
G.~W\"ackerle, S.~Appelt, and M.~Mehring.
%\newblock Spin-polarized noble gases: A playground for geometric quantum-phase
%  studies in magnetic resonance.
\newblock {\em Nucl. Instr. and Meth. in Phys. Res. A}, 402:464--472, 1998.

\bibitem{Pendlebury04_PRA}
J.~M.~Pendlebury {\it et al.}
%\newblock Geometric-phase-induced false electric fipole moment signals for
%  particles in traps.
\newblock {\em Phys. Rev. A}, 2004.

\bibitem{Lamoreaux05_PRA}
S.~K. Lamoreaux and R.~Goulb.
%\newblock Detailed discussion of a linear electric field frequency shift in
%  confined gases by a magnetic field gradient: Implications for neutron
%  electric dipole-moment measurements.
\newblock {\em Phys. Rev. A}, 2005.

\bibitem{stutz}
R.~Stutz and E.~Cornell.
\newblock {\em Bull. Am. Soc. Phys.}, 49:76, 2004.

\end{thebibliography}
\end{document}